\documentclass[smallabstract,smallcaptions]{dccpaper}
\usepackage[numbers,sort&compress]{natbib}
\renewcommand*{\citet}[2][]{{\cite[#1]{#2}}}

\usepackage{amsthm} 
\usepackage{thmtools}
\usepackage{amssymb}
\newtheorem{theorem}{Theorem}[section]
\usepackage{citesort}
\usepackage{hyperref}
\newenvironment{example}{\medskip\noindent\textbf{Example.}}{}
\newtheorem{definition}{Definition}

\newlength{\figurewidth}
\newlength{\smallfigurewidth}
\usepackage{graphicx}
\usepackage{xcolor}
\usepackage{booktabs}
\usepackage[capitalise]{cleveref}

\newcommand{\bsq}[1]{\lq{#1}\rq} 

\newcommand{\UnaryOperator}[2][]{%
	\ifx&#1&%
	\ensuremath{\mathop{}\mathopen{}#2\mathopen{}}%
	\else%
	\ensuremath{\mathop{}\mathopen{}#2\mathopen{}(#1)}%
	\fi%
}
\newcommand{\Oh}[1]{\UnaryOperator[#1]{\mathcal{O}}}
\newcommand{\oh}[1]{\UnaryOperator[#1]{o}}

\newcommand{\Om}[1]{\UnaryOperator[#1]{\mathup{\Omega}}}

\newcommand{\Ot}[1]{\UnaryOperator[#1]{\mathup{\Theta}}}
\DeclareMathAlphabet{\mathup}{OT1}{msb}{m}{n}

\newcommand*{\instancename}[1]{\ensuremath{\mathsf{#1}}} 
\newcommand*{\NSV}{\instancename{NSV}}
\newcommand*{\PSV}{\instancename{PSV}}
\newcommand*{\BWT}{\instancename{BWT}}
\newcommand*{\SA}{\instancename{SA}}
\newcommand*{\ISA}{\instancename{ISA}}
\newcommand*{\offT}{\ensuremath{\instancename{off}^{\instancename{T}}}}
\newcommand*{\off}{\instancename{off}}

\newcommand*{\functionname}[1]{{{\renewcommand{\rmdefault}{ptm}\fontfamily{ppl}\selectfont\textrm{\textup{#1}}}}} 
\newcommand*{\lce}{\functionname{lce}}
\newcommand*{\fnEnc}{\functionname{enc}}
\newcommand*{\LZ}{\functionname{LZ-text}}
\newcommand*{\HOLZ}{\functionname{HOLZ}}
\newcommand*{\Rev}[1]{\ensuremath{#1}^{\textup{R}}}

\begin{document}
\title{\large \textbf{HOLZ: High-Order Entropy Encoding of Lempel--Ziv Factor Distances}}
%
%
\author{
Dominik K\"oppl$^{\ast}$, Gonzalo Navarro$^{\dag}$, and Nicola Prezza$^{\ddag}$\\[0.5em]
{\small\begin{minipage}{\linewidth}\begin{center}
    \begin{tabular}{ccc}
		$^{\ast}$ M\&D Data Center& $^{\dag}$Dept. of Computer Science & $^{\ddag}$DAIS \\
    TMDU & University of Chile & Ca' Foscari University \\
    Tokyo, Japan & Santiago, Chile &  Venice, Italy\\
\url{koeppl.dsc@tmd.ac.jp} & \url{gnavarro@dcc.uchile.cl} & \url{nicola.prezza@unive.it}\\ \\
    \end{tabular}
    \end{center}
    \end{minipage}}
}
%
%
%
\maketitle              
\thispagestyle{empty}
\begin{abstract}
We propose a new representation of the offsets of the Lempel--Ziv (LZ) factorization
based on the co-lexicographic order of the processed prefixes.
The selected offsets tend to approach the $k$-th order empirical entropy.
Our evaluations show that this choice of offsets is superior to 
the rightmost LZ parsing and the bit-optimal LZ parsing on datasets with small high-order entropy.
\end{abstract}

\section{Introduction}
The Lempel--Ziv (LZ) factorization~\cite{lempel76lz}
is one of the most popular methods for lossless data compression. It builds on the idea of {\em factorization}, that is, splitting the text $T$ into {\em factors}, each being the longest string that appears before in $T$, and replacing each factor by a reference to its preceding occurrence (called its {\em source}). 

Most popular compression schemes such as \texttt{zip} or \texttt{gzip} use a variant called LZ77 \cite{ziv77lz}, which finds sources only within a sliding window~$w$.
Though this restriction simplifies compression and encoding, it misses repetitions with gaps larger than $|w|$, 
and thus compressing $k$ copies of a sufficiently long text~$T$ results in a compressed file being about $k$ times larger than the compressed file of $T$. 
Such long-spaced repetitions are common in highly-repetitive datasets like 
genomic collections of sequences from the same taxonomy group, or
from documents managed in a revision control system. Highly-repetitive datasets are among the fastest-growing ones in recent decades, and LZ compression is one of the most effective tools to compress them \cite{navarro21indexingi}.
This is one of the main reasons why the original LZ factorization (i.e., without a window) moved into the spotlight of recent research.

Although LZ catches even distant repetitions, the actual encoding of the factorization is an issue.
Each factor is usually represented by its length and the distance to its source (called its {\em offset}).
While the lengths usually exhibit a geometric distribution favorable for universal encoders,
the offsets tend to approach a uniform distribution and their codes are long. 
Since the sources are not uniquely defined, different tie breaks have been exploited in order to improve compression,
notably the \emph{rightmost} parsing (i.e., choosing the closest source) and the \emph{bit-optimal} parsing~\cite{ferragina13bit} (i.e., optimizing the size of the encoded file instead of the number of factors).

All previous LZ encodings have in common that they encode the {\em text distance} to the source. In this paper we propose a variant that encodes the distances between the co-lexicographically sorted prefixes of the processed text, and argue that this choice for the offsets approaches the $k$-th order empirical entropy of the LZ factors. Our experiments show that this encoding is competitive even with the bit-optimal parsing~\cite{ferragina13bit}, performing particularly well on texts whose high-order entropy is low.


\section{LZ Factorization}
Let $T[1..n] \in \Sigma^n$ be a text of length $n$ whose characters are drawn from an integer alphabet $\Sigma = [0..\sigma-1]$ with $\sigma = n^{\Oh{1}}$.
The LZ factorization is a partitioning of the text~$T$ into factors~$F_1 \cdots F_z$
such that every factor~$F_x$ is equal to the longest substring that starts before~$F_x$ in the text (its source).
Here, we imagine the text being prefixed by the characters of~$\Sigma$ in non-positive positions, 
i.e. $T[-c] = c$ for each $c \in \Sigma$.
Further, we compute the factorization on $T[1..n]$ while taking the prefixes of all suffixes starting in $T[-\sigma..p-1]$ under consideration for previous occurrences of a factor starting at $T[p]$.
In this setting, we always find a factor of length at least one.
Hence, each factor~$F_x$ can be encoded as a pair $(\offT_x, \ell_x)$, 
where $\ell_x = |F_x|$ is the length of~$F_x$, and $\offT_x$ is the \emph{textual offset} of~$F_x$, that is, the distance of the starting positions of $F_x$ and its source.
Thus, the LZ factorization can be represented as a sequence of these integer pairs of offsets and lengths,
which we denote by $\LZ(T)$.
This representation is not unique with respect to the choice of the offsets
in case of a factor having multiple occurrences that start before it.
The so-called \emph{rightmost} parsing selects always the smallest possible offset as a tie break.

\begin{figure}[t]
    \centering{%
	\includegraphics[scale=0.9,page=1]{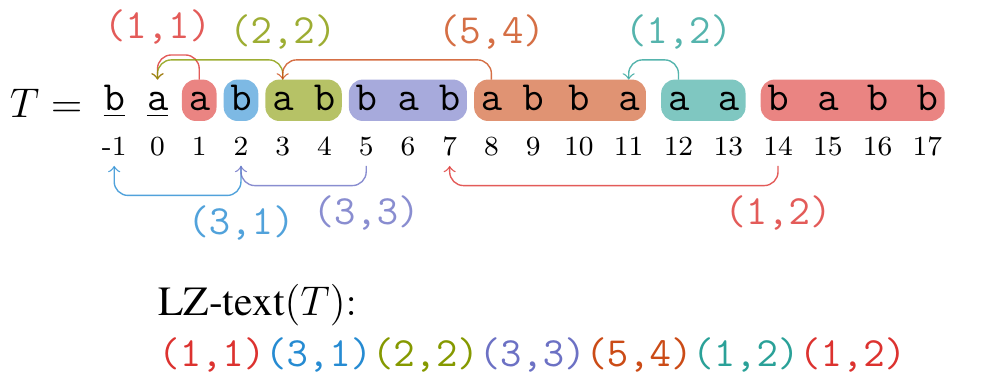}
    }
    \caption{LZ factorization of $T = \texttt{ababbababbaaababb}$ 
    and its $\protect\LZ(T)$ coding into pairs.}
    \label{figLZparse}
\end{figure}

\begin{example}
    Let $T = \texttt{ababbababbaaababb}$. Then the LZ factorization and its representation as pairs are given in \cref{figLZparse}. 
    Remember that we imagine the text as being prefixed by all characters of~$\Sigma$, being \texttt{ba} in this case;
    we only factorize~$T[1..|T|]$, that is, the non-underlined characters of \texttt{\underline{b}\underline{a}ababbababbaaababb}. \hfill $\Box$

\end{example}

\medskip
With the rightmost parsing it holds that, for a stationary ergodic source, 
the minimum number of bits required to write the number $\offT_x$, which is $\lceil \log_2(\offT_x+1) \rceil$,
approaches the zero-order entropy of $F_x$ in~$T$ (see~\cite{wyner94sliding} or \cite[Sec.~13.5.1]{cover12elements}). 
Intuitively, this happens because, being $\textrm{Pr}(F_x = S)$ the probability that the next LZ factor~$F_x$ equals the substring~$S$, 
on expectation we need to visit $\offT_x = 1/\textrm{Pr}(F_x = S)$ positions back (the log of which is $F_x$'s entropy) before encountering an occurrence of $S$. 
The fact that the factors are long ($\Ot{\log_\sigma n}$ characters on average) can be used to show that the size of $\LZ(T)$ approaches the $k$-th order entropy of~$T$, for small enough $k = \oh{\log_\sigma n}$.
In that light, we aim to reach the high-order entropy of the factors $F_x$, by means of changing the encoding of the sources.

\section{The HOLZ Encoding}\label{secHOLZ}
We now give a different representation of the sources of the LZ factors, 
and argue that this representation achieves high-order entropy.
Broadly speaking, we compute the offsets in co-lexicographic order\footnote{That is, the lexicographic order of the reversed strings.} of the text's prefixes, rather than in text order. 
We then compute a new offset~$\off_p$ instead of the textual offset $\offT_p$ used in $\LZ(T)$.

\begin{definition}
Let $T_i := T[-\sigma+1..i]$, for $i \in [-\sigma..n]$, be the prefixes of $T$, including the empty prefix 
$T_{-\sigma} = \epsilon$.
Assume we have computed the factorization of $T[1..p-1]$ for a text position~$p \in [1..n]$
and that a new factor~$F_x$ of length~$\ell_x$ starts at $T[p]$.

Let $T_{j_1} \prec T_{j_2} \prec \dots \prec T_{j_{p+\sigma}}$ be the 
prefixes~$T_{-\sigma},\ldots,T_{p-1}$ in co-lexicographic order.
Let $r_{p-1}$ be the position of $T_{p-1}$ in that order, that is, $j_{r_{p-1}} = p-1$, and let $t_{p-1}$ be the position closest to $r_{p-1}$ (i.e., minimizing $|r_{p-1} - t_{p-1}|$) satisfying $T[j_{t_{p-1}}+1..j_{t_{p-1}}+\ell_x] = F_x$.
Then we define $\off_x := r_{p-1} - t_{p-1}$; note $\off_x$ can be negative. 
We call $\HOLZ(T)$ the resulting encoding of pairs $(\off_x,\ell_x)$, where HO stands for high-order entropy.
\end{definition}

\begin{figure}
    \centering{%
	\includegraphics[scale=0.9,page=2]{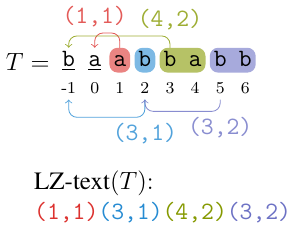}
	\includegraphics[scale=0.9,page=3]{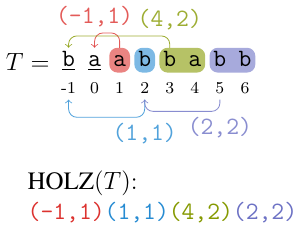}
    }
    \caption{LZ factorization of $T = \texttt{abbabb}$ with its $\protect\LZ$ (\emph{left}) and $\protect\HOLZ$ encoding (\emph{right}). Since the encodings only differ in the values of their offsets, they are equal in length.}
    \label{figLZHOparse}
\end{figure}

	\begin{table}
	\caption{Step-by-step computation of $\protect\HOLZ(T)$ with $T = \texttt{abbabb} = F_1 F_2 F_3 F_4$. Underlined characters represent the virtual text prefix containing all alphabet's characters.}
	\label{tabLZHOexample}
	    \centering
		\scriptsize
		\vspace{1mm}
	    \begin{minipage}[b]{0.45\textwidth}
		\begin{tabular}{lcr|ll}
		&&sorted& text to & \\
		&&prefixes& their right & \\\hline
			$T[-1..-2]$ & = & \texttt{$\epsilon$ }                & \texttt{~\underline b\underline aabbabb} & 1\\
			$T[-1..0]$  & = & \texttt{~\underline b\underline a } & \texttt{~abbabb}                          & $2 = r_0$\\
			$T[-1..-1]$ & = & \texttt{~\underline b }             & \texttt{~\underline aabbabb}              & $3=t_0$\\
		\end{tabular}
		\\
		\centering
		Computing $F_1 = T[1]$.
	\vspace{1em}
        \end{minipage}
	\hfill
	    \begin{minipage}[b]{0.45\textwidth}
		\begin{tabular}{lcr|ll}
		&&sorted& text to & \\
		&&prefixes& their right & \\\hline
			$T[-1..-2]$ & = & \texttt{$\epsilon$ }                  & \texttt{~\underline b\underline aabbabb} & $1=t_1$\\
			$T[-1..1]$  & = & \texttt{\ \underline b\underline aa } & \texttt{~bbabb}                           & $2=r_1$\\
			$T[-1..0]$  & = & \texttt{\ \underline b\underline a }  & \texttt{~abbabb}                          & 3\\
			$T[-1..-1]$ & = & \texttt{\ \underline b }              & \texttt{~\underline aabbabb}              & 4\\
		\end{tabular}
		\\
		\centering
		Computing $F_2 = T[2]$.
	\vspace{1em}
        \end{minipage}
 	    \begin{minipage}[b]{0.45\textwidth}
 		\begin{tabular}{lcr|ll}
		&&sorted& text to& \\
		&&prefixes& their right & \\\hline
			$T[-1..-2]$ & = & \texttt{$\epsilon$ }                   & \texttt{~\underline b\underline aabbabb} & $1=t_2$\\
			$T[-1..1]$  & = & \texttt{~\underline b\underline aa }  & \texttt{~bbabb}                           & $2$\\
			$T[-1..0]$  & = & \texttt{~\underline b\underline a }   & \texttt{~abbabb}                          & 3\\
			$T[-1..-1]$ & = & \texttt{~\underline b }               & \texttt{~\underline aabbabb}              & 4\\
			$T[-1..2]$  & = & \texttt{~\underline b\underline aab } & \texttt{~babb}                            & $5 = r_2$\\
		\end{tabular}	    
		\\
		\centering
		Computing $F_3 = T[3..4]$.
        \end{minipage}
	\hfill
 	    \begin{minipage}[b]{0.45\textwidth}
 		\begin{tabular}{lcr|ll}
		&&sorted&text to& \\
		&&prefixes&their right& \\\hline
			$T[-1..-2]$ & = & \texttt{$\epsilon$ }                    & \texttt{~\underline b\underline aabbabb} & 1\\
			$T[-1..1]$  & = & \texttt{~\underline b\underline aa }    & \texttt{~bbabb}                           & $2=t_4$\\
			$T[-1..0]$  & = & \texttt{~\underline b\underline a }     & \texttt{~abbabb}                          & 3\\
			$T[-1..4]$  & = & \texttt{~\underline b\underline aabba } & \texttt{~bb}                              & $4=r_4$\\
			$T[-1..-1]$ & = & \texttt{~\underline b }                 & \texttt{~\underline aabbabb}              & 5\\
			$T[-1..2]$  & = & \texttt{~\underline b\underline aab }   & \texttt{~babb}                            & 6\\
			$T[-1..3]$  & = & \texttt{~\underline b\underline aabb }  & \texttt{~abb}                             & 7
		\end{tabular}	    
		\\
		\centering
		Computing $F_4 = T[5..6]$.
        \end{minipage}       
	\end{table}

\begin{example}
    We present a complete step-by-step factorization of
    the small sample string $T = \texttt{abbabb}$. The
    $\HOLZ(T)$ factorization is given in \cref{figLZHOparse}, and
    \Cref{tabLZHOexample} depicts the four factorization steps; remember that we do not factorize the added prefix~$T[-\sigma+1..0]$.
    A detailed walk-through follows:
	\begin{enumerate}
	    \item For the first factor $F_1=\texttt{a}$ starting at $p=1$, the text prefix starting with $F_x$ and being the closest to the $r_0$-th co-lexicographically smallest prefix, has rank $t_0=3$, 
	    	thus $F_1$'s length and offset are $\ell_1 = 1$ and $\off_1 = r_0 - t_0 = 2-3 = -1$, respectively.
	    	We then represent $F_1 = T[1] = \texttt{a}$ by the pair $(-1,1)$.
	        \item Next, we update the table of the sorted prefixes, obtaining the order shown on the top-right table. 
		    The next factor, starting at position $p=2$, is $F_2=\texttt{b}$, so $\ell_2 = 1$ and $\off_2 = r_1-t_1 = 2-1 = 1$. 
	    	The second pair is thus $(1,1)$.
	        \item We update the table of the sorted prefixes, obtaining the order shown on the bottom-left table. 
		    This time, $p=3$, $F_3=\texttt{ba}$, $\ell_3 = 2$, and $\off_3 = r_2-t_2 = 5-1 = 4$. 
		    The third pair is thus $(4,2)$. 
	        \item We update the table of the sorted prefixes, obtaining the order shown on the bottom-right table; the final pair is $(2,2)$. 
	\end{enumerate}
Thus, we obtained $\HOLZ(T)$ = \texttt{(-1,1) (1,1) (4,2) (2,2)}.
\hfill $\Box$
\end{example}

\SubSection{Towards High-Order Entropy}




Only for the purpose of formalizing this idea, let us define a variant of HOLZ, $\HOLZ^k(T)$, which precedes $T$ with a (virtual) de Bruijn sequence of order $k+1$, so that every string of length $k+1$ appears in $T[-\sigma^{k+1}-k+2..0]$ (i.e., classical $\HOLZ(T)$ is $\HOLZ^0(T)$). 
We modify the LZ factorization so that $F_x$, starting at $T[p]$ and preceded by the string $S_x$ of length $k$, will be the longest prefix of $T[p..]$ such that $S_x \cdot F_x$ appears in $T$ starting before position $p-k$. The resulting factorization $T=F_1F_2\ldots$ has more factors than the LZ factorization, but in exchange, the offsets of the factors $F_x$ are encoded within their $k$-th order (empirical) entropy. 
Let $\#S$ be the frequency of substring $S$ in $T$.  Assuming that the occurrences of $S_x F_x$ distribute uniformly among the occurrences of $S_x$ in every prefix of $T$, the distance $|\off_x|$ between two consecutive sorted prefixes of $T$ suffixed by $S_x$ and followed by $F_x$ is in expectation $\mathbb{E}(|\off_x|) \le \#S_x / \#S_x F_x$. 
Then, $\mathbb{E}(\log_2 |\off_x|) \le \log_2 \mathbb{E}(|\off_x|) \le \log_2 (\#S_x / \#S_x F_x)$ and the total expected size of the encoded offsets is 
$$ \mathbb{E}\left(\sum_x \log_2 |\off_x| \right) ~~=~~ \sum_x \mathbb{E}(\log_2 |\off_x|) ~~\le~~ \sum_x \log_2 \frac{\#S_x}{\#S_x F_x}.$$

In the empirical-entropy sense (i.e., interpreting probabilities as relative frequencies in $T$), the definition of high-order entropy we can reach is restricted to the factors we produce. Interpreting conditional probability as following in the text, this is 
$$ H_k = \sum_x \log_2 \frac{1}{\textrm{Pr}(F_x|S_x)} = \sum_x \log_2 \frac{\textrm{Pr}(S_x)}{\textrm{Pr}(S_xF_x)} = \sum_x \log_2 \frac{\#S_x}{\#S_xF_x}.
$$

That is, the expected length of our encoding is bounded by the $k$-th order empirical entropy of the factors. This is also the $k$-th order empirical entropy of the text if we assume that the factors start at random text positions. 



Recall that, the longer $k$, the shorter the phrases, so there is an optimum for a likely small value of $k$. While this optimum may not be reached by HOLZ (which always chooses the longest phrase), it is reached by the bit-optimal variant of HOLZ that we describe in the next section, {\em simultaneously} for every $k$. 

Our experimental results validate our synthetic analysis, in the sense that $\HOLZ(T)$ performs better than $\LZ(T)$ on texts with lower $k$-th order entropy, for small $k$. 




\SubSection{Algorithmic Aspects}\label{secAlgorithm}

For an efficient computation of a factor~$F_x$ starting at~$T[p]$, 
we need to maintain the prefixes $T_{-\sigma},\ldots,T_{p-1}$ sorted in co-lexicographic order 
such that we can quickly find the ranks~$r_{p-1}$ of $T_{p-1}$ and the rank~$t_{p-1}$ of the starting position of an occurrence of $F_x$.
In what follows, we describe an efficient solution based on dynamic strings. 

Our algorithms in this paper work on the word RAM model with a word size of $\Om{\lg n}$ bits.
We use the following data structures:
\SA{} denotes the suffix array~\cite{manber93sa} of $T$,
such that $\SA[i]$ stores the starting position of the $i$-th lexicograhically smallest suffix in $T$. $\ISA{}$ is its inverse, $\ISA[\SA[i]]=i$ for all $i$.
The Burrows-Wheeler transform~\BWT{} of $T$ is defined by $\BWT[i] = T[\SA[i]-1]$ for $\SA[i] > 1$ and $\BWT[i] = T[n]$ for $\SA[i] = 1$.

We maintain a dynamic wavelet tree on the reverse of~$T[-\sigma..p-1]$.
This framework was already used \citet{policriti18lz77} to compute $\LZ(T)$ with the dynamic wavelet tree.
If we represent this wavelet tree with the dynamic string of \citet{munro15compressed}, 
their algorithm runs in $\Oh{n \log n/\log\log n}$ time using $nH_k + \oh{n\log\sigma}$ bits.

\begin{theorem}[\cite{munro15compressed}]\label{thmDynamicString}
	A dynamic string~$S[1..n]$ over an alphabet with size $\sigma$
	can be represented in $n H_k + \oh{n \log \sigma}$ bits 
	while supporting the queries access, rank, and select, as well as insertions or deletions of characters, in $\Oh{\log / \log \log n}$ time.
\end{theorem}

%
The text is not counted in this space; it can be given online in a streaming fashion
The main idea can be described as follows:
Let $R_i := \Rev{T_i}\texttt{\$}$.
Given that we want to compute a factor~$F_x$ starting at text position~$T[p]$, 
we interpret $T[p..]$ as the reverse of a pattern~$P$ that we want to search in $\BWT_{R_{p-1}}$ with the backward search steps of the FM-index~\cite{ferragina00fmindex}.
The backward search takes the last character of~$P$ being $T[p]$ as the initial search range in $\BWT_{R_{p-1}}$, 
updates $\BWT_{R_{p-1}}$ with the next character $T[p]$ to $\BWT_{R_{p}}$ and recurses on computing the range for $P[1..2] = \Rev{T[p..p+1]}$.
The recursion ends at the step before the range becomes empty.
In that case, all \BWT{} positions in the range correspond to occurrences of the factor~$F_x$ starting before~$F_x$ in $T$.
We choose the one being the closest (in co-lexicographic order) to the co-lexicographic rank of the current text prefix, and compute with $|F_x|$ forward traversals in the \BWT{} the original position in $\BWT_{R_{p-1}}$,
which gives us $t'_{p-1}$.
The position $r'_{p-1}$ is where $\texttt{\$}$ is stored in $\BWT_{R_{p-1}}$.
We wrote $t'_{p-1}$ and $r'_{p-1}$ instead of $t_{p-1}$ and $r_{p-1}$ since these positions are based on
$\BWT_{R_{p+|F_x|-1}}$ instead of $\BWT_{R_{p-1}}$.
We can calculate the actual offset $r_{p-1} - t_{p-1}$ by $r'_{p-1} - t'_{p-1}$ if we know the number of
newly inserted positions between $r'_{p-1}$ and $t'_{p-1}$ during the steps when we
turned $\BWT_{R_{p-1}}$ into $\BWT_{R_{p+|F_x|-1}}$.
For that, we additionally maintain a dynamic bit vector that marks the entries we inserted into
$\BWT_{R_{p-1}}$ to obtain $\BWT_{R_{p+|F_x|-1}}$
(alternatively, we can store the positions in a list).
We conclude that we can compute \HOLZ{} in $n(1+H_k) + \oh{n\log\sigma}$ bits of space and $\Oh{n\log n/\log\log n}$ time by using the data structure of \cref{thmDynamicString}.


%


\section{The Bit-Optimal HOLZ}
Ferragina et al.
\citet{ferragina13bit} studied a generalization of the Lempel--Ziv parsing in the sense that they considered for each text position all possible
factor candidates (not just the longest ones), optimizing for the representation minimizing a fixed encoding of the integers (e.g. Elias-$\delta$).
In other words, in their setting we are free to choose both the offset and 
factor lengths, thus 
effectively choosing among all possible unidirectional macro-schemes~\cite{storer82lzss}.
Within their framework, LZ can be understood as a greedy factorization that locally always chooses the longest factor among all candidates.
This factorization is optimal with respect to the number of computed factors,
but not when 
measuring the \emph{bit-size} of the  factors compressed by the chosen encoding for the integers, in general.
Given a universal code~$\fnEnc$ for the integers, a \emph{bit-optimal} parsing has the least number of bits 
among all unidirectional parsings using~$\fnEnc$ to encode their pairs of lengths and offsets.
In the setting of textual offsets, 
\citet{ferragina13bit} proposed an algorithm computing the bit-optimal LZ factorization in \Oh{n \lg n} time with \Oh{n} words of space, provided that the code~$\fnEnc$ transforms an integer in the range $[1..n]$ to a bit string of length $\Oh{\lg n}$.
In the following, we take this restriction of $\fnEnc$ as granted, as it reflects common encoders like Elias-$\gamma$.

\begin{figure}[h!]
    \centering{%
	\includegraphics[scale=1.3]{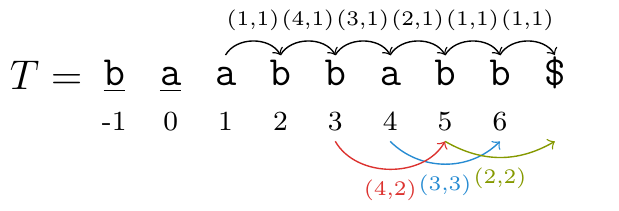}
    }
    \caption{%
	Graph of the factor candidates for $\protect\LZ(T)$. 
	Every path from node~$1$ to node~$n+1$ gives us a sequence of pairs that can be used alternatively to $\protect\LZ(T)$, which is obtained by always taking the locally longest edge.
	Although it is guaranteed that the path for $\protect\LZ(T)$ has the least number of edges, the compressed representation of the edge labels does not lead to the best compression in general. 
	Using Elias-$\gamma$ as our encoder~$\protect\fnEnc$ with 
	$|\protect\fnEnc(x)| = 1$ for $x=1$, 
	$|\protect\fnEnc(x)| = 2$ for $x\in\{2,3\}$, and
	$|\protect\fnEnc(x)| = 3$ for $x\in [4..7]$,
	the compressed size of $\protect\LZ(T)$ taking the red and the green arc is 
	$1+1 + 3+1 + 3+2 + 2 + 2 = 15$ bits.
	If we exchange the red and green arc with one blue arc and two black arcs,
	we obtain
	$1+1 + 3+1 + 1+1 +2+2 + 1+1 = 14$ bits.
    }
    \label{figBitOptimalExample}
\end{figure}

\SubSection{Factor Graph}
All possible unidirectional parsings using textual offsets can be represented by the following weighted directed acyclic graph: 
This graph has $n+1$ nodes, where the $i$-th node~$v_i$ corresponds to text position~$i \in [1..n]$, and the $(n+1)$-st node corresponds to the end of the text, which we can symbolize by adding an artificial character $\texttt{\$}$ to $T$ that is omitted in the factorization, see~\cref{figBitOptimalExample}.
An arc connecting node~$v_i$ with a node $v_{i+\ell-1}$ 
corresponds to a candidate factor starting at position~$i$ with length~$\ell$.
A \emph{candidate factor} is a pair~$(j,\ell)$ of position and length with $j < i$ and $T[j..j+\ell-1] = T[i..i+\ell-1]$. 
The weight of the arc~$(v_i, v_{i+\ell-1})$ corresponding to $(j,\ell)$ is the cost of encoding its respective factor,
$|\fnEnc(j)|+|\fnEnc(\ell)|$ if $|\fnEnc(x)|$ denotes the length of the binary output of $\fnEnc(x)$.
By construction, there are no arcs from a node~$i$ to a node~$j$ with $j < i$,
but there is always at least one arc from node~$i$ to a node~$j$ with $i < j$.
That is because we have at least one factor candidate starting at position~$i$
since we can always refer to a character in $T[-\sigma..0]$.
Hence, node~$1$ is connected with node~$n+1$.
Let the \emph{length} of an arc be the length of its corresponding factor, that is, if an arc connects node~$i$ with node~$j$, then its length is $j-i$.
We can then obtain the classic LZ factorization by following the maximum-length arcs starting from node~$1$, 
and we obtain the bit-optimal parsing by computing the (weighted) shortest path.

While we have $n$ nodes, we can have \Oh{n^2} arcs since the set of candidate factors for a text position~$i$ is
$\{ (j, \ell) : j < i \textup{~and~} T[j..j+\ell-1] = T[i..i+\ell-1] \}$, and this set can have a cardinality of \Oh{n}.
Theorem \citet[Thm.~5.3]{ferragina13bit} solves this issue by showing that it suffices to only consider so-called maximal arcs.
An arc~$(v_i, v_j)$ corresponding to a factor candidate~$(\off, \ell)$ is called \emph{maximal}~\cite[Def.~3.4]{farruggia19bicriteria}
if either there is no arc~$(v_i, v_{j+1})$, or such an arc corresponds to a factor candidate~$(\off', \ell')$
with $\fnEnc(\off)+\fnEnc(\ell) < \fnEnc(\off')+\fnEnc(\ell')$.
Moreover, there exists a shortest path from $v_1$ to $v_n$ that only consists of maximal arcs.
Since $|\fnEnc(\cdot)| \in \Oh{\lg n}$, we can divide all maximal arcs spawning from a node~$v_i$ into \Oh{\lg n} cost classes
such that each cost class is either empty or has exactly one maximal arc.
Hence, the pruned graph has just $\Oh{n \lg n}$ arcs.
Finding the shortest path can be done with Dijkstra's algorithm running in \Oh{n \lg n} time with Fibonacci heaps.
The computed shortest path is a sequence of factors that we encode in our final output.
Storing the complete graph would take \Oh{n \lg n} words. 
To get the space down to \Oh{n} words, the idea \citet[Cor.~5.5]{ferragina13bit} is to compute the arcs online 
by partitioning, for each cost class~$k \in [1..\Oh{\lg n}]$, the text into blocks, and process each of the blocks in batch.

\SubSection{The Bit-Optimal \protect\HOLZ{}}
By exchanging the definition of $\off$ in the factor candidates, we compute a variation of \HOLZ{} that is bit-optimal.
The major problem is finding the maximal arcs efficiently.
Like before, we maintain the dynamic BWT of the reversed processed text.
But now we also maintain a dynamic wavelet tree mapping suffix ranks of the processed reversed text 
to suffix ranks of~$T$.
Then we can compute for each cost class of arcs spawning from~$v_i$ the maximal arc by searching 
the suffix ranks closest to the suffix rank of~$T[i..]$ within the interval of suffix ranks of the reversed processed text.

\newcommand*{\DyWa}{\instancename{DyWa}}
\newcommand*{\CurRsaIdx}{\ensuremath{p_\texttt{\$}}}
\newcommand*{\maxSAp}{\ensuremath{\instancename{maxSA}^p}}
\newcommand*{\predSAp}{\ensuremath{\instancename{predSA}^p}}
\newcommand*{\succSAp}{\ensuremath{\instancename{succSA}^p}}
\newcommand*{\predRSAp}{\ensuremath{\instancename{predRSA}^p}}
\newcommand*{\succRSAp}{\ensuremath{\instancename{succRSA}^p}}
\newcommand*{\maxRSAp}{\ensuremath{\instancename{maxRSA}^p}}

In what follows, we explain our algorithm computing the factor candidates corresponding to all maximal arcs.
Let again $T[-\sigma+1..n]$ be the text with all distinct characters prepended, 
and let $\Rev{T}$ denote its reverse.
In a preprocessing step, we build \SA{} and \ISA{} on $T[-\sigma+1..n]$.
Like before, we scan the text~$T[1..n]$ from left to right.
Let $p$ be the text position where are currently processing, such that $T[1..p-1]$ has already been processed.
Let $R_{p-1} := \Rev{T_{p-1}} \texttt{\$}$ denote the string whose BWT we maintain in $\BWT_{R_{p-1}}$.
Further, let $\DyWa$ be a dynamic wavelet tree mapping $\ISA_{R_{p-1}}[i]$ to $\ISA_T[i]$ for each $i \in [1..|R_{p-1}|]$.
Before starting the factorization, we index/process $T[-\sigma-1..0]^R$ with~$\BWT$ and~$\DyWa$.
Since $\ISA_{R_{p-1}}$ is not necessarily a prefix of $\ISA_{R_{p}}$,
for adding a new entry to \DyWa{} when processing a position~$p \in [-\sigma+1..n]$, 
we first need to know $\ISA_{R_{p}}[1]$, i.e., the rank of the suffix in $R_p$ corresponding to $T[p]$ in $\BWT_{R_{p-1}}$.
Luckily, this is given by the position of $\texttt{\$}$ in $\BWT_{R_{p-1}}$.
Let \CurRsaIdx{} denote this position in $\BWT_{R_{p-1}}$, 
and suppose that we have processed $T[1..p-1]$.
Now, for each of the ranges $I_1 = [1..\CurRsaIdx-1]$ and $I_2 = [\CurRsaIdx+1..\sigma+p-1]$,
let $\predRSAp_j$ and $\succRSAp_j$ be values in the range $I_j$ that are mapped via \DyWa{}
to the smallest value larger than $\ISA[p]$ and the largest value smaller than $\ISA[p]$, respectively.
Let us call these mapped values $\predSAp_j$ and $\succSAp_j$.
%
Let $\maxSAp_j$ be the one that has a longer LCE value with $\ISA[p]$, i.e., we compare $\lce(T[\SA[\predSAp_j]..], T[p..])$ with $\lce(T[\SA[\succSAp_j..], T[p..])$.
Finally, let $\maxSAp$ be the one among $\maxSAp_1$ and $\maxSAp_2$ having the largest LCE~$\ell$ with $p$,
and let $\maxRSAp$ be its corresponding position in $\BWT_{R_{p-1}}$.
This already defines the factor candidate with offset $\CurRsaIdx - \maxRSAp$ and the longest length~$\ell$ among all factor candidates starting at $T[p]$.
Next, we compute the factor candidates with smaller lengths but less costly offsets.
For that, we partition $[\maxRSAp_1..\CurRsaIdx-1]$ into cost classes, i.e., 
intervals of ranks whose differences to $\CurRsaIdx$ need the same amount of bits when compressed via a universal coder.
This gives $\Oh{\lg n}$ intervals, and for each interval we perform the same query as above to retrieve a $\overline{\maxSAp}$ value
with $\lce(T[\SA[\overline{\maxSAp}]..], T[p..]) \le \lce(T[\SA[\maxSAp_1]..], T[p..])$.
We start with the interval with the shortest costs, and keep track of the maximal LCE value computed up so far.
For each candidate interval, if its computed maximal LCE value is not larger than the already computed LCE value, then we discard it
since it would produce a factor with the same length but a higher cost for the offset.
We cut $I_1$ at $\maxRSAp_1$ since this gives us the maximum LCE value in the entire range, so going further does not help us in discovering an even longer candidate factor.
We process $I_2$ with $\maxRSAp_2$ symmetrically.
In total, we obtain for each text position \Oh{\lg n} factor candidates, which we collect in a list per text position;
the summed size of these lists is $\Oh{(\sigma + n) \lg n}$.
Each pair~$(\ell, \off)$ in the list of position~$p$ consists of its length~$\ell$ and the offset~$\off$ being
the difference between $\CurRsaIdx$ and its respective position~$j$ in $\BWT_{R_{p-1}}$.
The cost for encoding this pair is~$|\fnEnc(\ell)| + |\fnEnc(|\off|)|$ incremented by one for a bit storing whether $j > \CurRsaIdx$ or not.
The computed pairs determine the edges of the weighted directed acyclic graph explained above.

%
%

\paragraph{Complexities}
The wavelet tree~\DyWa{} stores $\Oh{n}$ integers in the range $[1..n]$ dynamically.
For each of the \Oh{\lg n} levels, a rank or select query can be performed in \Oh{\lg n} time,
and thus updates or queries like range queries take \Oh{\lg^2 n} time.
For each text position, we perform one update and $\Oh{\lg n}$ range queries,
accumulating to \Oh{n \lg^3 n} time overall for the operations on~\DyWa{}.
This is also the bottleneck of our bit-optimal algorithm using the \HOLZ{} encoding.
The wavelet tree representing the dynamic BWT of the reversed processed text works exactly as the previous algorithm in Section~\ref{secAlgorithm},
and the produced graph needs \Oh{n \lg n} time for processing.
Summing up, our algorithm to compute the bit-optimal \HOLZ{} encoding runs in \Oh{n \lg^3 n} time and uses \Oh{n \lg n} words of space. The space could be improved to \Oh{n} words using the same techniques discussed in \citet{ferragina13bit}.

\paragraph{Decompression}
The decompression works in both variants (\HOLZ{} or its bit-optimal variant) in the same way.
We maintain a dynamic BWT on the reversed processed text; When processing a factor~$F$ starting at text position~$p$ and encoded with pair~$(\off,\ell)$,
we know that BWT position $t_{p-1} = \CurRsaIdx - \off$ corresponds to the starting text position of the factor's source. It is then sufficient to extract $\ell$ characters from that position, by navigating the BWT using LF queries. 
At each query, we also extend the BWT with the new extracted character to take into account the possibility that the source overlaps $F$. 
Overall, using the dynamic string data structure of \cref{thmDynamicString}, the decompression algorithm runs in $O(n\log n/\log\log n)$ time and uses $nH_k + o(n\log\sigma)$ bits of space (excluding the input, which however can be streamed). 

\section{Experiments}
For our experiments, we focus on the Canterbury and on the Pizza\&Chili corpus.
For the latter, we took 20 MB prefixes of the data sets.
See \cref{tableCanterbury,tablePizzaChili} for some characteristics of the used datasets.
The datasets \textsc{kennedy.xls}, \textsc{ptt5}, and \textsc{sum} contain \bsq{0} bytes,
which is a prohibited value for some used tools like suffix array construction algorithms.
In a precomputation step for these files, 
we escaped each \bsq{0} byte with the byte pair \bsq{254} \bsq{1} and each former occurrence of \bsq{254} with the pair \bsq{254} \bsq{254}.

\begin{table}[t]
    \caption{20 MB prefixes of the Pizza\&Chili corpus datasets. $H_k$ denotes the $k$-th order empirical entropy. $z$ is the length of $\protect\LZ(T)$, and $r$ is the number of runs of $\protect\BWT$ built upon the respective dataset.}
    \label{tablePizzaChili}
    \centerline{%
	\begin{tabular}{l*{10}{r}}
	    \toprule
dataset                & $\sigma$ & $z$      & $r$     & $H_0$ & $H_1$ & $H_2$ & $H_3$ & $H_4$ \\
	    \midrule
\textsc{cere         } & 5        & 8492391  & 1060062 & 2.20  & 1.79  & 1.79  & 1.78  & 1.78 \\
\textsc{coreutils    } & 235      & 3010281  & 910043  & 5.45  & 4.09  & 2.84  & 1.85  & 1.31 \\
\textsc{dblp.xml     } & 96       & 3042484  & 834349  & 5.22  & 3.26  & 1.94  & 1.26  & 0.89 \\
\textsc{dna          } & 14       & 12706704 & 1567099 & 1.98  & 1.93  & 1.92  & 1.92  & 1.91 \\
\textsc{e.coli       } & 11       & 8834711  & 1146785 & 1.99  & 1.98  & 1.96  & 1.95  & 1.94 \\
\textsc{english      } & 143      & 5478169  & 1277729 & 4.53  & 3.58  & 2.89  & 2.33  & 1.94 \\
\textsc{influenza    } & 15       & 876677   & 210728  & 1.97  & 1.93  & 1.93  & 1.92  & 1.91 \\
\textsc{kernel       } & 160      & 1667038  & 488869  & 5.38  & 4.00  & 2.87  & 1.98  & 1.47 \\
\textsc{para         } & 5        & 8254129  & 1028222 & 2.17  & 1.83  & 1.83  & 1.82  & 1.82 \\
\textsc{pitches      } & 129      & 10407645 & 2816494 & 5.62  & 4.85  & 4.28  & 3.50  & 2.18 \\
\textsc{proteins     } & 25       & 8499596  & 1958634 & 4.20  & 4.17  & 4.07  & 3.71  & 2.97 \\
\textsc{sources      } & 111      & 4878823  & 1361892 & 5.52  & 4.06  & 2.98  & 2.13  & 1.60 \\
\textsc{worldleaders } & 89       & 408308   & 129146  & 4.09  & 2.46  & 1.74  & 1.16  & 0.73 \\
 	    \bottomrule
	\end{tabular}
    }
\end{table}

\begin{table}[t]
    \caption{Datasets from the Canterbury corpus; $n$ is the number of text characters. See \cref{tablePizzaChili} for a description of the other columns.}
    \label{tableCanterbury}
    \centerline{%
	\begin{tabular}{l*{9}{r}}
	    \toprule
dataset & $n$ & $\sigma$ & $z$ & $r$ & $H_0$ & $H_1$ & $H_2$ & $H_3$ & $H_4$ \\                                                                               
	    \midrule
\textsc{alice29.txt }& 152089  & 74  & 66903  & 22897  & 4.56 & 3.41 & 2.48 & 1.77 & 1.32 \\
\textsc{asyoulik.txt}& 125179  & 68  & 62366  & 21634  & 4.80 & 3.41 & 2.53 & 1.89 & 1.37 \\
\textsc{cp.html     }& 24603   & 86  & 9199   & 4577   & 5.22 & 3.46 & 1.73 & 0.77 & 0.44 \\
\textsc{fields.c    }& 11150   & 90  & 3411   & 1868   & 5.00 & 2.95 & 1.47 & 0.86 & 0.62 \\
\textsc{grammar.lsp }& 3721    & 76  & 1345   & 853    & 4.63 & 2.80 & 1.28 & 0.67 & 0.44 \\
\textsc{kennedy.xls }& 1486290 & 255 & 219649 & 145097 & 3.13 & 2.04 & 1.76 & 1.19 & 1.12 \\
\textsc{lcet10.txt  }& 426754  & 84  & 165711 & 52594  & 4.66 & 3.49 & 2.61 & 1.83 & 1.37 \\
\textsc{plrabn12.txt}& 481861  & 81  & 243559 & 72622  & 4.53 & 3.36 & 2.71 & 2.13 & 1.72 \\
\textsc{ptt5        }& 961861  & 158 & 65867  & 25331  & 1.60 & 0.47 & 0.39 & 0.31 & 0.26 \\
\textsc{sum         }& 50503   & 254 & 13544  & 7826   & 4.76 & 2.52 & 1.61 & 1.15 & 0.90 \\
\textsc{xargs.1     }& 4227    & 74  & 2010   & 1172   & 4.90 & 3.19 & 1.55 & 0.72 & 0.42 \\
	    \bottomrule
	\end{tabular}
    }
\end{table}

\begin{figure}
    \centering{%
	\includegraphics[width=0.8\linewidth,page=1]{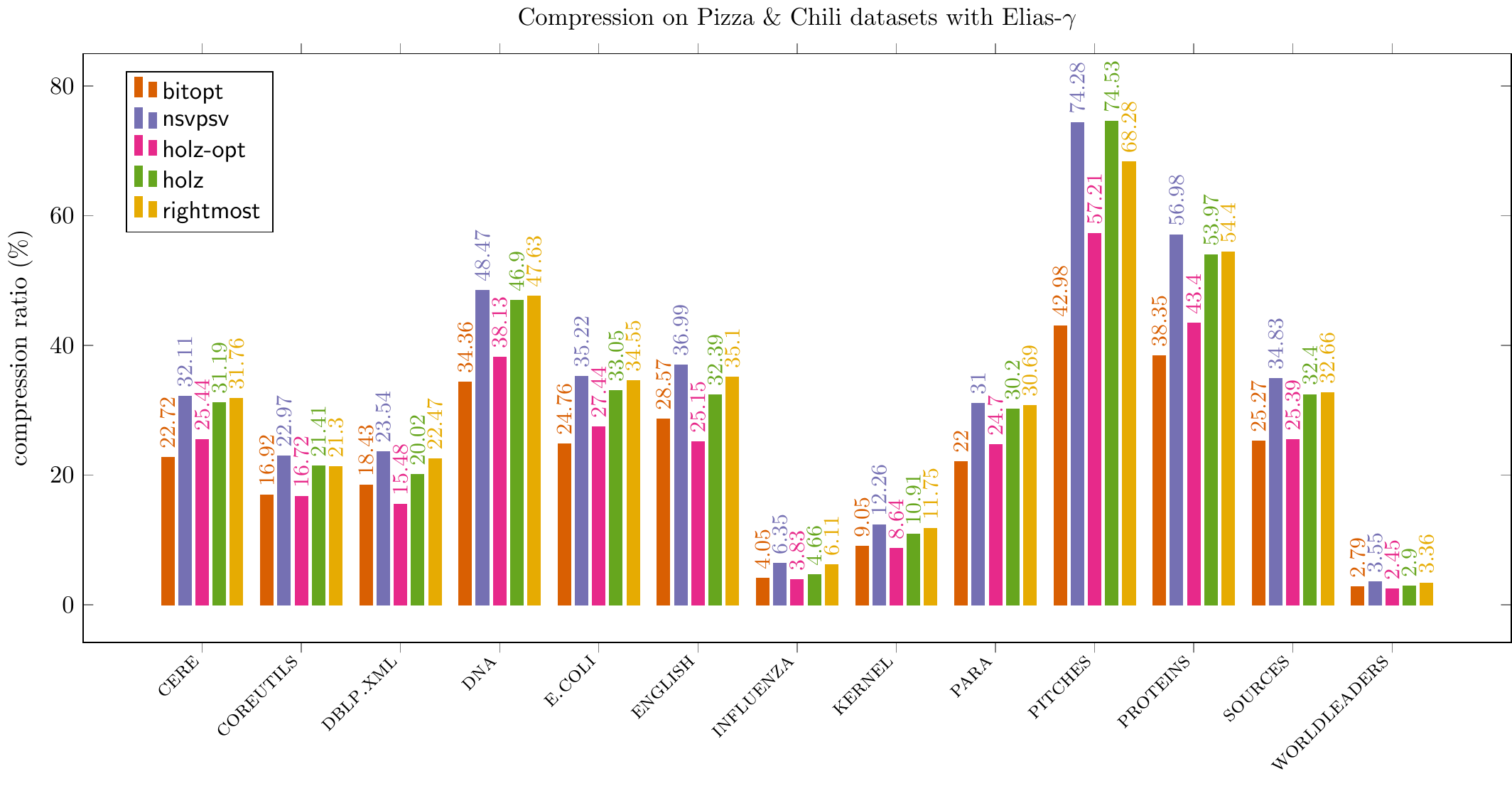}
	\includegraphics[width=0.8\linewidth,page=2]{plot/plot.pdf}
    }
    \caption{Compression ratios of different encodings and parsings studied in this paper on the Pizza\&Chili corpus.}
    \label{figCompressionRatioPC}
\end{figure}

\begin{figure}
    \centering{%
	\includegraphics[width=0.8\linewidth,page=3]{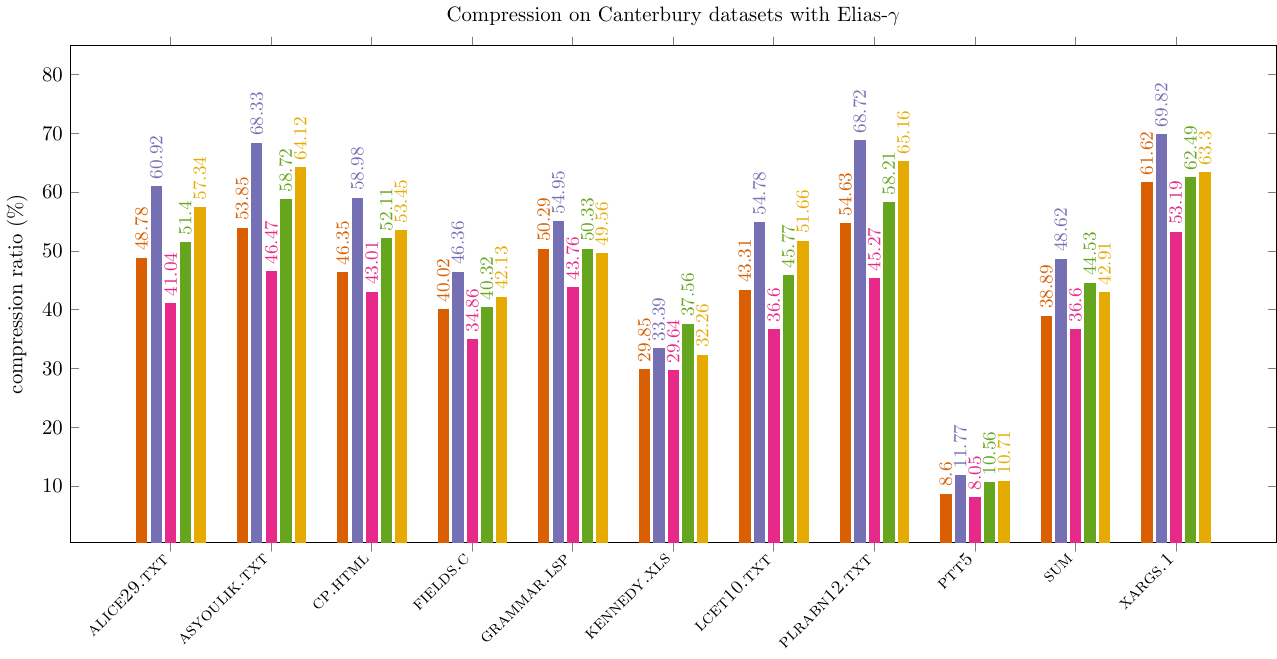}
	\includegraphics[page=5]{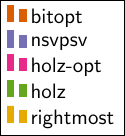}
	\includegraphics[width=0.8\linewidth,page=4]{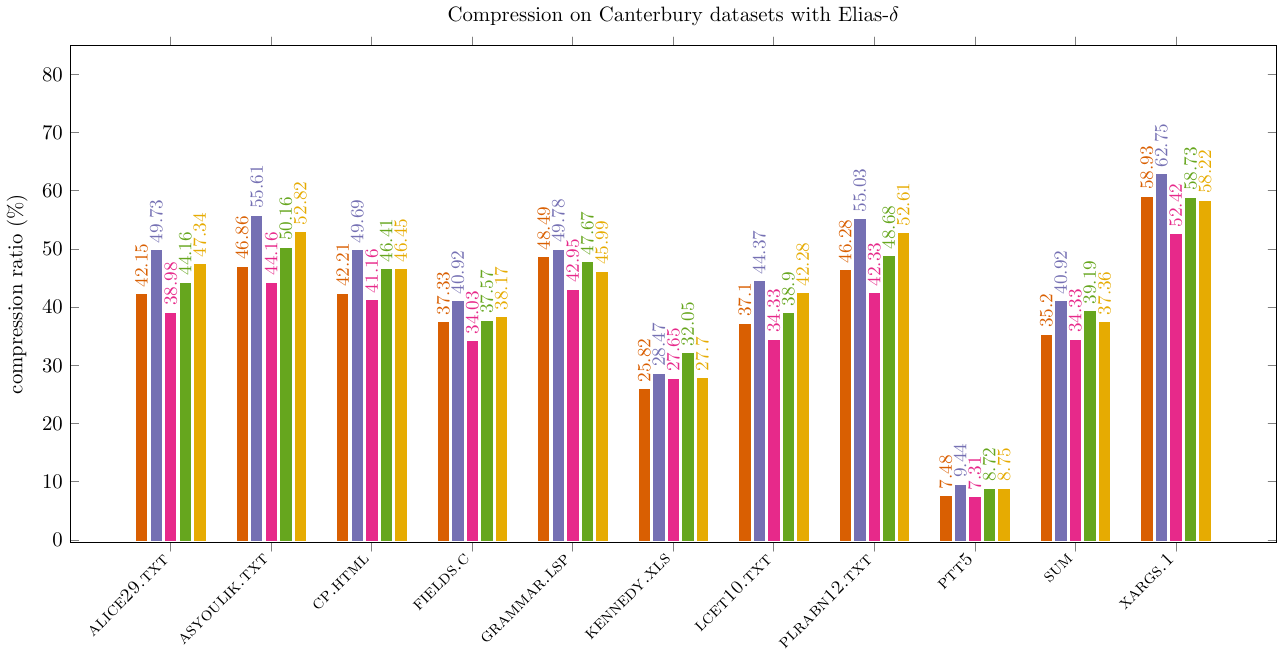}
    }
	\caption{Compression ratios of different encodings and parsings studied in this paper on the Canterbury corpus. 
	}
    \label{figCompressionRatioCanterbury}
\end{figure}

For comparison, we used $\LZ(T)$ and the bit-optimal implementation of \citet{farruggia19bicriteria}, referred to as \textsf{bitopt} in the following.
For the former, we remember that the choice of the offsets can be ambiguous.
Here, we select two different codings that break ties in a systematic manner:
The rightmost parsing introduced in the introduction, and the output of an algorithm~\cite{ohlebusch11lz} computing LZ with next-smaller value (NSV) and previous-smaller value (PSV) arrays built on the suffix array.
The PSV and NSV arrays, \PSV{} and \NSV{}, store at their $i$-th entry the largest index~$j$ smaller than $i$ (resp.\ smallest index~$j$ larger than $i$) with $\SA[j] < i$.
Given a starting position~$p$ of a factor, candidates for the starting positions of its previous occurrences 
are stored at the entries $\NSV[\ISA[p]]$ and $\PSV[\ISA[p]]$ of \SA{}.
We take the one that has the larger LCE with~$p$, say $r$, and output the offset~$p-r$.
Although the compressed output is at least that of the rightmost parsing, this algorithm runs in linear time, 
whereas we are unaware of an algorithm computing the rightmost parsing in linear time -- the currently best algorithm needs $\Oh{n + n\log \sigma / \sqrt{\log n}}$ time~\cite{belazzougui16rightmost}.
We call these two specializations \textsf{rightmost} and \textsf{nsvpsv}.
We selected Elias-$\gamma$ or Elias-$\delta$ encoding as the function \fnEnc{}, and 
present the measured compression ratios in \cref{figCompressionRatioPC,figCompressionRatioCanterbury}.
We write \textsf{holz} and \textsf{holz-opt} for our presented encoding $\HOLZ(T)$ and its bit-optimal variant, respectively.

We observe that the bit-optimal implementation uses a different variant of the LZ factorization 
that does not use the imaginary prefix $T[-\sigma..0]$ for references.
Instead, it introduces \emph{literal factors} that catch the leftmost occurrences of each character appearing in~$T$.
Their idea is to store a literal factor~$S$ by its length~$|S|$ encoded in 32-bits, followed by the characters byte-encoded.
In the worst case, they pay $40\sigma$ bits.
This can pose an additional overhead for tiny files such as \textsc{grammar.lsp}, where $40\sigma$ bits are roughly 20\% of their output size.
However, it becomes negligible with files like \textsc{cp.html}, where less than 4\% of the output size can be accounted for the literal factors.


\paragraph{Discussion}

The overall trend that we observe is that our encoding scheme HOLZ performs better than LZ on datasets characterized by a small high-order entropy. 
More in detail, when focusing on Elias-$\delta$ encoding (results are similar for Elias-$\gamma$), \texttt{holz-bitopt} compresses better than \texttt{bitopt} on all 8 datasets having $H_4 \leq 1$, and on 11 over 14 datasets with $H_4 \leq 1.5$. In general, \texttt{holz-bitopt} performed no worse (strictly better in all cases except one) than \texttt{bitopt} on 14 over 24 datasets. The trend is similar when comparing the versions of the algorithms that always maximize factor length: \texttt{holz} and \texttt{rightmost}. 
These results support the intuition that HOLZ is able to exploit high-order entropy, improving when it is small enough the compression ratio of the offsets with respect to LZ.

\section{Open Problems}
We wonder whether there is a connection between our proposed encoding \HOLZ{} and the Burrows-Wheeler transform (BWT) combined with move-to-front (MTF) coding,
which achieves the $k$-th order entropy of~$T$ with $k = \Ot{\log_\sigma n}$.
Applying MTF to BWT basically produces a list of pointers, where each pointer refers to the closest previous occurrence of a character whose context is given by the lexicographic order of its succeeding suffix.
The difference is that this technique works on characters rather than on factors.
%
%
Also, we would like to find string families where the compressed output of $\HOLZ(T)$ is asymptotically smaller than $\LZ(T)$, or vice-versa.

Our current implementation using dynamic wavelet trees is quite slow.
Alternatively, for the encoding process we could use a static BWT and a dynamic prefix sum structure to mark visited prefixes in co-lexicographic order,
which should be faster than a dynamic BWT\@. 
A more promising alternative would be to not use dynamic structures.
We are confident that a good heuristic by hashing prefixes according to some locality-sensitive hash function (sensitive to the co-lexicographic order) will find matches much faster. 
Note that using the context of length $k$ has the additional benefit to reduce the search space (compared to standard LZ), therefore the algorithm could be much faster and it could be easier to find potential matches. 

\subsection*{Acknowledgements}
This research was funded by JSPS KAKENHI with grant numbers JP21H05847 and JP21K17701, and by Fondecy Grant 1-200038 (Chile).
We are grateful to github user MitI\_7 for some guidance regarding the dynamic wavelet tree implementation.

\bibliographystyle{IEEEbib}
\bibliography{main}
\end{document}